\documentclass[
onecolumn,
groupedaddress,
 aps,
prd,
]{revtex4-2}
\usepackage[colorinlistoftodos]{todonotes}
\usepackage{comment}
\usepackage{ulem}
\usepackage{hyperref}
\usepackage{graphicx}
\usepackage{subcaption}
\usepackage{array}
\usepackage{amssymb}
\usepackage{amsmath}
\usepackage{tabularx}
\usepackage{mathtools}
\usepackage{setspace}
\usepackage{float}
\usepackage{array}
\usepackage{graphicx}% Include figure files
\usepackage{dcolumn}% Align table columns on decimal point
\usepackage{bm}% bold math
\usepackage{comment}
\usepackage[colorinlistoftodos]{todonotes}
\usepackage{comment}
\usepackage{ulem}
\usepackage{hyperref}
\usepackage{graphicx}
\usepackage{dcolumn}
\usepackage{caption}
\usepackage{subcaption}
\usepackage{array}
\newcolumntype{P}[1]{>{\centering\arraybackslash}p{#1}}
\usepackage{bm,color}
\usepackage{amssymb}
\usepackage{amsmath}
\usepackage{comment}
\usepackage{makecell,tabularx,multirow}
\usepackage{mathtools}
\usepackage{times}
\usepackage{bigints}
\usepackage{enumitem}
\usepackage{float}
\usepackage{booktabs}
\usepackage{graphicx}% Include figure files
\usepackage{dcolumn}% Align table columns on decimal point
\usepackage{bm}% bold math
\usepackage{comment}
\usepackage{dcolumn}% Align table columns on decimal point
\usepackage{bm}% bold math
\usepackage{amsmath}
\usepackage{amsfonts}
\usepackage{siunitx}
\usepackage[english]{babel}
\usepackage{amsfonts}
\usepackage{braket,slashed,bm}
\usepackage{natbib}
\usepackage{multirow}
\usepackage{hyperref}
\usepackage[noabbrev]{cleveref}
\usepackage{lipsum}

\hypersetup{
    colorlinks=true,
    linkcolor=blue,
    filecolor=blue,      
    urlcolor=blue,
    anchorcolor=blue,
    citecolor=blue
}
\def\beq{\begin{equation}}
\def\eeq{\end{equation}}
\def\barr{\begin{array}}
\def\earr{\end{array}}
\def\dis{\displaystyle}

\newcommand{\be}{\begin{equation}}
\newcommand{\ee}{\end{equation}}
\newcommand{\bea}{\begin{eqnarray}}
\newcommand{\eea}{\end{eqnarray}}

\newcommand{\bi}{\begin{itemize}}
	\newcommand{\ei}{\end{itemize}}

\begin{document}
\title{Constraint on cosmological constant in generalized Skryme-teleparallel system}% Force line breaks with \\

\author{Krishnanand K. Nair}
 \email{krishnanandk@vt.edu}
\affiliation{
Department of Physics, Virginia Tech, Blacksburg, VA, 24061, USA\\
}

\author{Mathew Thomas Arun}%
 \email{mathewthomas@iisertvm.ac.in}
\affiliation{
School of Physics, Indian Institute of Science Education and Research, Thiruvananthapuram 695551, India\\
}

\begin{abstract}

The Einstein-Skyrme system is understood to defy the "no hair" conjecture by possessing black-hole solutions with fractional baryon number outside the event horizon. In this article, we extend the study of the Skyrme system to teleparallel gravity framework. We consider two scenarios, the Teleparallel Equivalent of General Relativity (TEGR) and generalized teleparallel gravity $f(T)$. In our analysis, we compute the fractional baryon number beyond the black-hole horizon and its correlation with the cosmological constant ($\Lambda$). In the TEGR context, where $f(T) = -T - 2\Lambda$, the results match with the Einstein-Skyrme model, assuming a positive $\Lambda$. More interestingly, in generalized teleparallel gravity scenario, defined by $f(T) = -T - \tau T^2 - 2\Lambda$, we show that the existence of a solution demands that not only must $\Lambda$ be positive but has to lie in a range, $\Lambda_{min} < \Lambda < \Lambda_{max}$. While the upper bound depends inversely on $\tau$, the lower bound is a linear function of it. Hence, in the limiting case with generalized teleparallel gravity converging towards TEGR ($\tau \rightarrow 0$), the constraints on the cosmological constant relax to the Einstein Skryme system ($\Lambda_{min}$ approaches zero and $\Lambda_{max}$ becomes unbounded). On the other hand, in f(T) gravity, vanishing cosmological constant solution is found only if the lower bound on the energy of the soliton is very large.
\end{abstract}
%\keywords{Suggested keywords}%Use showkeys class option if keyword
                              %display desired
 \maketitle

%\tableofcontents

\section{Introduction}
General Relativity (GR) has been remarkably effective in explaining a wide range of cosmological events, yet some phenomena still require the introduction of exotic matter. The nature of these hypothetical fields remains a mystery, particularly because of their elusive nature in collider experiments that explore beyond the Standard Model of particle physics. This uncertainty prompts a debate about whether to modify the matter sector or gravity itself. In recent times, there has been a growing interest in developing alternatives to General Relativity, with teleparallel gravity, especially, experiencing a revival. This approach uses a torsion-based Weitzenböck connection, in contrast to the curvature-based Levi-Civita connection used in GR~\cite{Cai:2015emx, Krssak:2018ywd, Krssak:2019, Aldrovandi:2013wha, Maluf:2013gaa, DeAndrade:2000sf, Pereira:2013qza, Fontanini:2018krt, Bahamonde:2021gfp}. A key aspect of teleparallel gravity is the Teleparallel Equivalence of General Relativity (TEGR), which parallels GR in certain aspects of its phenomenology. In its simplest form, the Ricci scalar from Einstein-Hilbert action is linked to the torsion scalar and a corresponding boundary term, ensuring that the solutions from general relativity remain valid in this new framework~\cite{deAndrade:2000kr}. Teleparallel scalar-tensor theories, or $f(T)$ theories, however, offer distinct solutions to field equations~\cite{Bahamonde:2022lvh, Rodrigues:2013ifa}. Over the past decade, extensive research covering various topics like bouncing universes~\cite{delaCruz-Dombriz:2018nvt, Nair:2021hnm}, black holes~\cite{bh1,bh2,bh3,bh4,bh5,bh6,bh7,bh8,bh9,bh10,bh11,bh12,bh13,bh14,bh15}, cosmological inflation~\cite{Awad:2017ign,RezaeiAkbarieh:2018ijw, Chakrabortty:2021tdl}, and gravitational waves~\cite{Capozziello:2019msc,Capozziello:2020vil,Li:2020xjt,Bahamonde:2021dqn,Najera:2021afa,gw1} has significantly heightened interest in this field. Like GR, exploring the generalizations of TEGR is also proving to be a valuable pursuit.

The Skyrme model is one such example from GR that rose to fame for having an intriguing feature that defies the "{\it no hair}" conjecture. According to this conjecture, all other information, such as global charges, are lost during the gravitational collapse that creates a black hole, which is entirely determined by its mass, electric charge, and angular momentum. For black holes with classical skyrmion hair, this has been contested and falsified \cite{Shiiki:2005pb, Dvali:2016mur}. In our study, we expand the Skyrme system to the frameworks of TEGR ($f(T) = -T-2\Lambda$) and generalized teleparallel gravity ($f(T) = -T - \tau T^2 -2\Lambda$), and we compute the fractional baryon number and its dependence on the cosmological constant $\Lambda$.

Skyrme is a stable solution in non-linear sigma model which have had miraculous success in modelling the low-energy dynamics of light pions. These particles arise as the bound states known as baryons which are charged under the $U(1)_V$ group that is preserved on chiral symmetry breaking, $U(3)_L \times U(3)_R \to U(1)_V \times SU(3)_V$. However, the chiral Lagrangian defining pions do not explain these baryons. It was soon discovered that the non-linear sigma model does possess a topology to support soliton solutions with the conserved current,
\begin{equation}
    B^\mu = \frac{1}{24\sqrt{-g}\pi^2}\epsilon^{\mu \nu \rho \sigma} Tr\Big(U^{\dagger}(\partial_\nu U)U^{\dagger}(\partial_\rho U)U^{\dagger}(\partial_\sigma U)\Big) \ \nonumber ,
\end{equation}
where, $U=e^{\frac{2i}{\kappa} \pi^a T^a}$ with $\pi^a$ representing pions and $\kappa$ their decay constant.

The simplest solution, with two derivative chiral Lagrangian, is not static by simple scaling arguments, but require higher derivative terms leading to the well-known Skyrme model given by,
\begin{equation}
    \mathcal{L}=\frac{\kappa^2}{4}Tr(\partial^\mu U^{\dagger}\partial_\mu U) + \frac{1}{32 e^2} Tr([U^{\dagger}\partial^\mu U, U^{\dagger}\partial^\nu U][U^{\dagger}\partial_\mu U, U^{\dagger}\partial_\nu U]) \nonumber \ ,
\end{equation}
where $\kappa$ is the pion decay constant and $e^2$ is a dimensionless coupling. This coupling determines the scale of the Skyrmion.  This Lagrangian leads to the winding number, $B=\int \sqrt{-g}d^3x B^0$, which is bounded from below and the soliton configurations with non-trivial winding numbers are identified as baryons. And the energy of the soliton can be seen to be bounded below by,
\begin{equation}
    E \geq \frac{6 \pi^2 \kappa}{e}|B| \ ,
    \label{eq:solitonenergy}
\end{equation}
where $|B|$ is the baryon number. 
Such topologically stable solutions of this model could have interesting effects in the early epochs of the Universe. Moreover, this model will couple to gravity and studying this can be insightful to understand the gravitational effects of systems with baryonic charges.

Though this subject has been studied in the context of General Relativity, here, we extend it to teleparallel gravity formalism. The reason to do this 2 fold. One, even though TEGR and GR are equivalent up to a boundary term, generalized teleparallel gravity is distinct. Moreover, objects with torsion do find a natural place in teleparallel gravity framework. Second, to describe a spinor field, like the baryon, GR is inadequate and need to move to theory of Einstein-Cartan. On the other hand, defining the torsion through Weitzenb\"{o}ck connection is much more elegant.
The building blocks in TEGR formalism are the four tangent space vectors erected on the space-time manifold, known as "tetrad fields" given by $h^a_\mu$, where $\mu$, $a$ = $\{1,\ 2,\ 3,\ 4\}$. From this the space-time metric can be constructed as
\begin{equation}
   g_{\mu \nu}= \eta_{ab}h^a{}_\mu h^b{}_\mu \ .
    \label{mp}
\end{equation}
 Here, $g_{\mu \nu}$ is the metric tensor, $\eta_{ab}$ is the Minkowski metric of the tangent space, and $h^a{}_\mu$ are the tetrads. In this paper, the Greek indices ($\mu$, $\nu$) refers to the space-time manifold and the Latin indices (a, b) to the local Minkowski tangent space $T_xM$.
These tetrads are used to establish the teleparallel equivalent of Levi-Civita connection, called the Weitzenb\"{o}ck connection $\Gamma_{\mu \nu}^{\rho}$ \cite{wb}, given by,
\begin{equation}
\Gamma^{\rho}{}_{\mu \nu}=h_{a}{}^{\rho} \partial_{\mu} h^{a}{ }_{\nu}
\label{wbconn}
\end{equation}
The calculation of the torsion tensor $T^a{}_{\mu \nu}$ \cite{Cai:2015emx} is thus made possible by the Weitzenb\"{o}ck connection. This tensor characterizes the torsion of spacetime, as given by,
\begin{equation}
T^{\rho}{}_{\mu \nu}=\Gamma^{\rho}{}_{\mu \nu}-\Gamma^{\rho}{}_{\nu\mu } \ .
\label{torsion}
\end{equation}
The Weitzenb\"{o}ck connection and Levi-Civita connection are related through the contorsion tensor ($K^{\rho}{}_{\mu \nu}$) as $\Gamma_{\mu \nu}^\rho - K^{\rho}{}_{\mu \nu} = \tilde{\Gamma}_{\mu \nu}^\rho$ where,
\begin{equation}
K^{\rho}{}_{\mu \nu}=\frac{1}{2}\left(T_{\mu}{ }^{\rho}{ }_{\nu}+T_{\nu}{ }^{\rho}{ }_{\mu}-T^{\rho}{}_{\mu \nu}\right) \ .
\label{contorsion}
\end{equation}
Further we define the dual torsion tensor as 
\begin{equation}
S^{\rho \mu \nu}= \frac{1}{2}\left[K^{\mu \nu \rho}-g^{\rho \nu} T^{\lambda \mu}{}_{\lambda}+g^{\rho \mu} T^{\lambda \nu}{}_{ \lambda}\right] \ .
\label{dualtorsion}
\end{equation}
Finally, the torsion scalar T, which is a quadratic function of torsion, becomes,
\begin{equation}
T= T_{\rho \mu \nu} S^{\rho \mu \nu}=T^{\rho}{}_{\mu \nu} T_{\rho}{ }^{\mu \nu} /2+T^{\rho}{}_{\mu \nu} T^{\nu \mu}{}_{\rho}-2 T^{\rho}{}_{\mu \rho} T^{\nu \mu}{ }_{\nu} \ .
\label{tscalar}
\end{equation}
Thus the teleparallel action now can be written interms of the Torsion scalar as,
\begin{equation}
      S_{\text{T}} = -\frac{1}{16\pi G}\int d^4x \ h_{det}\operatorname{T} 
      \nonumber
      \end{equation}
where $h_{det} = det(h^a_\mu) = \sqrt{-g}$.

Mathematically, the torsion scalar $T$ in teleparallel gravity is related to the Ricci scalar $\Tilde{R}$ in GR as \cite{RevModPhys.48.393, PhysRevD.19.3524}
\begin{equation}
    T \equiv -\Tilde{R} + B \ \nonumber,
\end{equation}
where $\Tilde{R}$ is the Ricci scalar and $B=2\Tilde{\nabla}_\mu(T^\nu{}_{\nu}{}^{\mu})$ is a total divergence term.
This TEGR Lagrangian can be readily extended to power law gravity ($f(T)$) \cite{PhysRevD.75.084031, PhysRevD.78.124019, PhysRevD.79.124019, PhysRevD.83.023508, Wei:2011aa, Atazadeh:2011aa}, analogous to how the Einstein-Hilbert action can be generalized to $f(R)$. The relationship between $T$ and $\Tilde{R}$ ensures that simplest teleparallel gravity and GR are equivalent, but $f(T)$ gravity can be quite different from $f(R)$ gravity, due to the presence of the boundary term which is studied extensively in literature~\cite{Cai:2015emx, Krssak:2018ywd, Krssak:2019}.

In this study, we investigate black hole solutions characterized by the presence or absence of baryonic charge ($B=0$ and $B \neq 0$) within the frameworks of TEGR and modified teleparallel gravity ($f(T)$). The subsequent section provides an overview of the teleparallel-Skyrme system. Section \eqref{sec:B=0} is dedicated to exploring the Skyrme solution and the corresponding adjustments to the metric in the vicinity of a black hole for a Skyrme model with $B=0$ baryonic charge. Additionally, in section \eqref{sec:Bneq0}, we delve into the analysis of solutions and the occurrence of fractional baryon charge in both TEGR and power law gravity contexts. We also analyze the constraints imposed by generalized teleparallel geometry on the cosmological constant. Finally, section \eqref{sec:results} offers a comprehensive conclusion and summary of our findings.

\section{Teleparallel-Skyrme system} 
\label{sec:tele-skyrme}
Similar to the Einstein's General Theory of Relativity, in teleparallel geometry, the minimal coupling prescription is given by\cite{deAndrade:1997cj,Nair:2021hnm},  \begin{equation}
\begin{aligned}
&\eta^{a b} \rightarrow g^{\mu \nu}=\eta^{a b} h_{a}^{\mu} h_{b}^{\nu} \\
&\partial_{a} \rightarrow \nabla_{\mu}=\partial_{\mu}- \Gamma_\mu \ ,
%\frac{i}{2}\Omega^{a b}{ }_{\mu} J_{a b} \ ,
\end{aligned}
\label{mcoupling}
\nonumber
\end{equation}
where $\Gamma_\mu $ is the Weitzen\"{o}ck connection.
%, $\Omega^{a b}{ }_{\mu}$ and $J_{ab}$  $\mathcal{D}_{\mu}$ are the Fock–Ivanenko Derivative Operator (FIDO) \cite{Fock1929, Nair:2021hnm}, and 
Thus, in the case of the Skyrme field expressed as $SU(2)$ group-valued scalar field U, the minimal coupling prescription becomes \cite{Nair:2021hnm}, 
\begin{equation}
    \partial_{a}U \rightarrow \nabla_{\mu}U \ \nonumber ,
\end{equation}
and, the Teleparallel-Skyrme action in four dimensions, $S=S_{\text{T}} + S_{\text{S}}$, is dictated by the generalized teleparallel gravity action ($S_{\text{T}}$) and the Skyrme action ($S_{\text{S}}$) given by
 \begin{eqnarray}
      S_{\text{T}} &=& \dis \frac{1}{16\pi G} \int d^4x \ h_{det} \operatorname{f(T)} \nonumber \\
S_{\text{S}} &=& \dis \int d^{4} x  h_{det} \left[\frac{\kappa^{2}}{4} \operatorname{Tr}\left(Q_{\mu} Q^{\mu}\right)+\frac{1}{32 e^{2}} \operatorname{Tr}\left(\left[Q_{\mu}, Q_{\nu}\right]\left[Q^{\mu}, Q^{\nu}\right]\right)\right] \nonumber \ \ \ ,
\label{eq:skyrme}
\end{eqnarray}
where, f(T) is a function of the torsion scalar $T$. In the action, $h_{det} = det(h^a_\mu) = \sqrt{-g}$ and G is the Newton's constant. In the Skyrme action, $Q_{\mu}=U^{-1}\partial_\mu U$ is the current in nonlinear sigma model. On varying this action w.r.t the tetrads $h^{a}_\mu$, we obtain the equation of motion \cite{Capozziello:2018qcp, Farrugia:2018gyz, Krssak:2015oua},
\begin{equation}
 M_\beta{}^\mu \equiv S_\beta{ }^{\alpha\mu } \partial_\alpha T f_{T T}(T)+\left[h_{det}^{-1} h^a{ }_\beta \partial_\alpha\left(h_{det}  h_a{ }^\sigma S_\sigma{ }^{\alpha \mu}\right)-T^\sigma{ }_{\nu \beta} S_\sigma{ }^{\mu \nu}\right] f_T(T)-\frac{1}{4} \delta_\beta^\mu f(T)=4 \pi G \mathcal{T}_\beta^\mu
 \label{eom1}
\end{equation}
\\
where $f_T=\frac{\partial f}{\partial T}$ and $f_{TT}=\frac{\partial^2 f}{\partial T^2}$. Here, $\mathcal{T}_{\mu \nu}$ represents the energy momentum tensor and, in terms of the matter Lagrangian density $\mathcal{L}_{M}$ becomes \cite{Bahamonde:2021gfp},
\begin{equation}
\mathcal{T}_\mu{}^\nu=\frac{h_{det}^{-1} h_{\mu}^{a}}{4}\left\{\frac{\partial \mathcal{L}_{M}}{\partial h_{\mu}^{a}}-\partial_{\alpha}\left[\frac{\partial \mathcal{L}_{M}}{\partial\left(\partial_{\alpha} h^{a}{ }_{\nu}\right)}\right]\right\} \
\nonumber.
\end{equation}
For the case of Skyrme field, this energy momentum tensor is computed to be,
\begin{align}
\mathcal{T}_{\mu \nu}= & -\frac{\kappa^2}{2} \operatorname{Tr}\Big(Q_{\mu} Q_{\nu}-\frac{1}{2} g_{\mu \nu} Q_{\alpha} Q^{\alpha}\Big)  -\frac{1}{8 e^{2}} \operatorname{Tr}\Big(g^{\alpha \beta}[Q_{\mu}, Q_{\alpha}][Q_{\nu}, Q_{\beta}]  -  \frac{1}{4} g_{\mu \nu}[Q_{\alpha}, Q_{\beta}][Q^{\alpha}, Q^{\beta}]\Big)
\label{emtensor}
\end{align}
To obtain the Skyrme equations of motions, varying the action, we get,
\begin{equation}
\nabla^{\mu}\left( Q_{\mu}+\frac{1}{4\kappa^{2} e^{2}}\left[Q^{\nu},\left[Q_{\mu}, Q_{\nu}\right]\right]\right)=0
\label{skyrmeeom}
\end{equation}
where $\nabla^{\mu}$ is the Fock Ivanenko Derivative \cite{Nair:2021hnm}. This equation of motion admits topological solutions where $U\to 1$ as $|x^\mu| \to \infty$. These static field configurations in the chiral Lagrangian is characterised by the homotopy class $\Pi_3(SU(2)) = \mathbb{Z}$, with the energy density bounded by the topological winding number (baryon number) $B \in \mathbb{Z}$. The topological current corresponding to the baryon current of the teleparallel-Skyrme system is given by \cite{Ioannidou:2006mg}
\begin{equation}
B^\mu=\frac{1}{h_{det}} \frac{1}{24 \pi^2} \epsilon^{\mu \nu \alpha \beta} \operatorname{Tr}\left(Q_\nu Q_\alpha Q_\beta\right)
\label{B-current}
\end{equation}
from which the baryon number B can be derived as
\begin{equation}
B=\int h_{det} B^0 d^3 x
\label{B}
\end{equation}
where $B^0$ is the the temporal component of the topological current.
To obtain the static solutions, let's first use the metric ansatz~\cite{Ioannidou:2006mg},
\begin{equation}
d s^{2}=-h(r) d t^{2}+\frac{1}{h(r)}\left(p_{1}(r) d r^{2}+p_{2}(r) r^{2} d \theta^{2}+p_3(r) r^{2} \sin ^{2} \theta d \varphi^{2}\right)
\label{eq:initialmetric}
\end{equation}
where $h(r)$, $p_1(r)$, $p_2(r)$, and $p_3(r)$ are functions of the coordinate $r$.\\
\textcolor{red}{}Analysis of the Skyrme system in teleparallel framework is new and has nontrivial solutions. The generalized hedgehog ansatz \cite{hedge1} has just recently made it possible to construct the first analytic gravitating Skyrmions and exact configurations of multi-Skyrmions \cite{hedge2, hedge3, hedge4, hedge5, hedge6, hedge7, hedge8, Ioannidou:2006mg, Ioannidou:2006nn}.
The general ansatz for the form of the $SU(2)-$valued Skyrme field U is given by \cite{hedge1},
\begin{equation}
U=\rho \cdot 1 + \pi^k \cdot t_k
\nonumber
\end{equation}
Here, $t_k$ are the $SU(2)$ generators, given by the Pauli matrices, 
\begin{equation}
    t_1=\begin{bmatrix}
  0 & -i\\ 
  i & 0
\end{bmatrix}\, \ \ \ t_2 =  \begin{bmatrix}
  0 & -1\\ 
  1 & 0
\end{bmatrix}\, \ \ \ t_3 =  \begin{bmatrix}
  -i & 0\\ 
  0 & i
\end{bmatrix}
\nonumber
\end{equation}
These matrices are Hermitian and traceless, and they satisfy the commutation relations $[t_i,t_j]=2i\epsilon_{ijk}t_k$, where $\epsilon_{ijk}$ is the totally antisymmetric Levi-Civita tensor. The index $k = 1, 2, 3$ corresponds to the $SU(2)$ group index, which is raised and lowered using the flat metric $\delta_{ij}$. The quartet of the field $(\rho, \pi_a)$ is restricted to the surface of the unit sphere, where the field is a map from compactified Euclidean coordinate space $S^3$ to the $SU(2)$ group space, satisfying $\rho^2 + \pi^k.\pi^k = 1$ \cite{Ioannidou:2006nn}. This constraint ensures that the Skyrme field is restricted to the surface of the unit sphere. It also implies that the Skyrme field has a topological charge, which is an integer-valued quantity that characterizes the topology of the field configuration. To specify the Skyrme field more explicitly, we use the anzats in \cite{Canfora:2013osa}, which takes the form,
\begin{equation}
U=1 \cos \gamma(r)+\widehat{n}^{k} t_{k} \sin \gamma(r) \ . 
\label{U}
\end{equation}
In above, we $\rho = \cos \gamma(r)$ and $\pi^k = \widehat{n}^{k} \sin \gamma(r)$, where $\gamma(r)$, and $\widehat{n}^{k}$ are functions of the coordinates $r$ and ($\theta$, $\varphi$) respectively, given by,
\begin{align}
\label{n1}
\widehat{n}^{1}= &\sin \theta  \cos \varphi \\
\label{n2}
\widehat{n}^{2}= &\sin \theta \sin \varphi\\ 
\label{n3}\widehat{n}^{3}= &\cos \theta
\end{align}
Thus, we can see that the Skyrme field $U$ is a matrix-valued function that depends on the radial coordinate $r$ and the unit vectors $\widehat{n}^k$. The ansatz in Eq. \eqref{U} respects the symmetry of the problem  and simplifies the analysis of Skyrme models. Substituting Eq. \eqref{n1}-\eqref{n3} in Eq. \eqref{U}, we get the expression of U as:
\begin{equation}
    U=\begin{bmatrix}
  \cos \gamma + i \cos \theta \sin \gamma  & \sin \theta \sin \gamma \left(i \cos \varphi + \sin \varphi\right) \\ 
  \sin \theta \sin \gamma \left(i \cos \varphi - \sin \varphi\right) &  \cos \gamma - i \cos \theta \sin \gamma
\end{bmatrix}
\label{U}
\end{equation}

Solving the components of Skyrme field and energy moment tensor of the Skymre field (given in Appendix \eqref{A1}), the vanishing condition (Eq.\eqref{Tconsv}) of $ \nabla_\mu \mathcal{T}^{\mu r}$ demand either $\gamma'=0$ or $\eta(r)=0$, while $\nabla_\mu \mathcal{T}^{\mu \theta}$ demands $p_2=p_3$. In order to satisfy the energy-momentum conservation in Eq. \eqref{Tconsv}, we need to impose the conditions,  
 \begin{align}
    p_2(r)=  p_3&(r)=m(r), \ \ \ \ \ \  p_1(r)=  l(r)
    \nonumber
\end{align}
where we rename $p_1(r)$ as $l(r)$ for convenience and we will solve for $\gamma(r)$ that satisfies $ \nabla_\mu \mathcal{T}^{\mu r}=0$.
 It is also important to note that equation Eq. \eqref{Tconsv} satisfies the energy momentum tensor components given in Eq. \eqref{T11} if the metric functions $h$, $p_1$, $p_2$, $p_3$, and $\gamma$ do not depend on $\theta$ and $\varphi$. Therefore, we can satisfy all the equations in Eq. \eqref{Tconsv} only in the spherically symmetric situation when $p_2(r)=p_3(r)=m(r)$, and the Skyrme and metric functions depend only on the radial coordinate.
The metric, given in Eq.\eqref{eq:initialmetric}, now becomes,
\begin{equation}
d s^{2}=-h(r) d t^{2}+\frac{l(r)}{h(r)}d r^{2}+\frac{m(r)}{h(r)}\left( r^{2} d \theta^{2}+ r^{2} \sin ^{2} \theta d \varphi^{2}\right)
\label{metricfinal}
\end{equation}
The respective teleparallel equations of motion are solved in the Appendix \eqref{A2}. Now let's first attempt to solve the Skyrme field equation given in Eq. \eqref{skyrmeeom}. One can easily decompose the Skyrme field $Q_\mu$, given in Eq. \eqref{Q3},  in terms of the SU(2) generators $t_k$ as,
\begin{equation}
    Q_\mu=Q_\mu^kt_k \ ,
\end{equation}
where $Q_\mu^k$ are given by,
\begin{equation}\label{Qtk}
\begin{split}
    Q_t^k=&0\\
    Q_r^k= & \widehat{n}^{k}\gamma'\\
    Q_\theta^k= & \sin^2\gamma\delta^{rk} \varepsilon_{ijr} \widehat{n}^{i}\partial_\theta \widehat{n}^{j}+\frac{1}{2}\sin(2\gamma)\partial_\theta \widehat{n}^{k}\\
     Q_\varphi^k= & \sin^2\gamma \delta^{rk} \varepsilon_{ijr} \widehat{n}^{i}\partial_\phi \widehat{n}^{j}+\frac{1}{2}\sin(2\gamma)\partial_\phi \widehat{n}^{k}
\end{split}
\end{equation}
Using Eqs. \eqref{Qtk}, we can easily obtain the divergence of $Q_\mu$, given in Eq. \eqref{Qdiv} as,
\begin{equation}
   \nabla^\mu Q_\mu = \frac{1}{2} \left(-\frac{h' \gamma'}{l}+\frac{h' \gamma'}{h^2}+\frac{h
   \left(r \left(2 r l m \gamma''+\gamma' \left(2 l \left(r m'+2
   m\right)-r m l'\right)\right)-2 l^2 \sin (2 \gamma)\right)}{r^2 l^2
   m}\right)\widehat{n}^{k}t_k
   \label{Qdiv}
\end{equation}
To compute the second term in Eq.\eqref{skyrmeeom}, we use,
\begin{equation}
\left[Q^{\nu},\left[Q_{\mu}, Q_{\nu}\right]\right] = W_\mu^k t_k,
\nonumber
\end{equation}
where $W_\mu^k$ are given as,
\begin{equation}\label{Wtk}
\begin{split}
    W_t^k=&0\\
    W_r^k=&8Q_r^k\frac{h(r)}{r^2m(r)}\sin^2 \gamma \\
W_\theta^k=&4Q_\theta^k\left({\gamma'}^2+\frac{h(r)}{m(r)r^2}\sin^2 \gamma\right)\\
W_\varphi^k=&4Q_\varphi^k\left({\gamma'}^2+\frac{h(r)}{m(r)r^2}\sin^2 \gamma\right)
\end{split}
\end{equation}
It is straightforward to obtain this divergence term as,
\begin{align}
    \nabla^\mu\left[Q^{\nu},\left[Q_{\mu}, Q_{\nu}\right]\right] = & 4 h \Bigg(\frac{h' \gamma' \sin ^2(\gamma)}{lr^2 m}+\frac{h' \gamma' \sin ^2(\gamma)}{h^2r^2 m}+h \Bigg(\frac{\sin (\gamma) \Big(2 l \left(\gamma'' \sin
   (\gamma)+\gamma'^2 \cos (\gamma)\right)-\gamma' \sin (\gamma) l'\Big)}{l^2r^2 m}\nonumber\\&-\frac{2 \sin ^3(\gamma) \cos (\gamma)}{r^2 r^2 m}\Bigg)\Bigg)\widehat{n}^{k}t_k
   \label{skyrmeeompart1}
\end{align}
After substituting Eq.\eqref{Qdiv} and Eq.\eqref{skyrmeeompart1} into Eq.\eqref{skyrmeeom} and simplifying, we obtain the equation of motion of the Skyrme field as, 
\begin{align}
  &  \frac{8}{e^2 \kappa ^2 r^2
   m}  \left(\frac{h' \gamma' \sin ^2(\gamma)}{l}+\frac{h' \gamma' \sin ^2(\gamma)}{h^2}+h \left(\frac{\sin (\gamma) \left(2 l \left(\gamma''
   \sin (\gamma)+\gamma'^2 \cos (\gamma)\right)-\gamma' \sin (\gamma) l'\right)}{l^2}-\frac{2 \sin ^3(\gamma) \cos (\gamma)}{r^2 m}\right)\right)\nonumber\\&-\frac{h' \gamma'}{l}+\frac{h' \gamma'}{h^2}+\frac{h \left(r \left(2 r l m \gamma''+\gamma' \left(2 l \left(r m'+2 m\right)-r m l'\right)\right)-2 l^2
   \sin (2 \gamma)\right)}{r^2 l^2 m}=0
   \label{Skyrmeeomf}
\end{align}
The Skyrme field equation of motion is a second order differential equation of $\gamma(r)$, that includes the metric functions $h(r)$, $l(r)$, and $m(r)$. The Skyrme function $\gamma(r)$ is important in determining the topological charge (given in Eq. \eqref{Baryoncharge}), while the metric functions provide useful information on the curvature and geometry of the spacetime manifold coupled with the Skyrmion. As a result, a thorough understanding of these functions in teleparallel gravity is required in order to completely explain the behaviour of the Skyrme field and its interactions with the underlying spacetime metric in teleparallel gravity.
Further, the topological charge (Eq. \eqref{B}) for the Skyrme anzats given by the Eq. \eqref{U} reduces to \cite{Luckock:1986tr}
\begin{equation}
B=\left.\frac{1}{\pi}\left\{-\gamma(r)+\frac{1}{2} \sin [2 \gamma(r)]\right\}\right|_{r_h} ^{\infty}
\label{Baryoncharge}
\end{equation}
where $r_h$ is the event horizon radius of the black-hole background. Thus, the problem reduces to solving a single ordinary differential equation for the Skyrme function $\gamma(r)$ given in Eq. \eqref{Skyrmeeomf}.
\section{Case 1: Skyrmions with B=0 }
\label{sec:B=0}
In the context of Skyrme theory, the simplest nontrivial solution of the Skyrme equation Eq.\eqref{Skyrmeeomf} is given by,
\begin{equation}
    \gamma=\frac{\pi}{2}+N\pi
    \label{gammasol}
\end{equation}
where N is an integer. The energy momentum tensor now becomes,
\begin{equation}\label{emcase1-1}
\begin{split}
    \mathcal{T}^t{}_t = & \mathcal{T}^r_r=-\frac{1}{2 e^2 r^4 m(r)^2}\Big(h(r) \left(2 e^2 \kappa ^2 r^2 m(r)+h(r)\right)\Big)\\
 \mathcal{T}^\theta{}_\theta = & \mathcal{T}^\varphi{}_\varphi= \frac{h(r)^2}{2 e^2 r^4 m(r)^2} 
\end{split}
\end{equation}
\begin{figure}[H]
    \centering
    \includegraphics[scale=0.64]{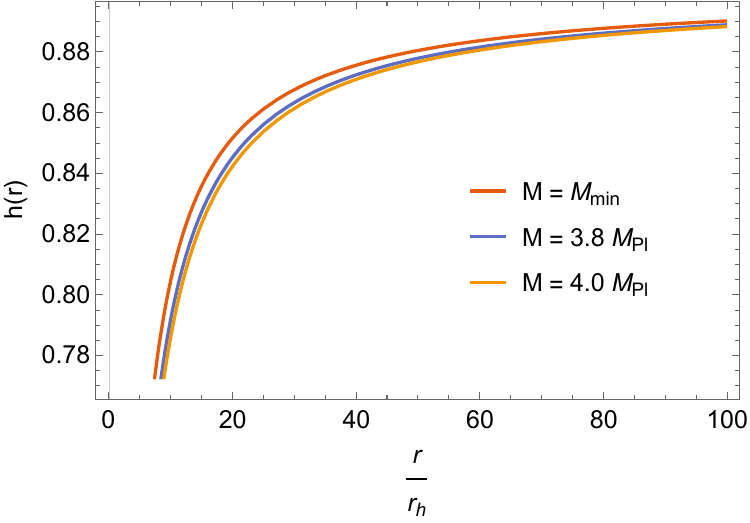}
    \caption{The metric function $h(r)$ is shown as a function of $r/r_h$ for different black hole masses $M$= $M_{min}$, $3.8 \ M_{Pl}$ and $4.0 \ M_{Pl}$.$M_{min}$ is computed using Eq.\eqref{Mmin} to be 0.36299 $M_{Pl}$. We have chosen the coupling constant $\alpha= 8\pi G\kappa^2 = 0.1$ , $G=1 M_{Pl}^{-2}$, $e=1$ and $\Lambda= 10^{-120} M_{Pl}^2$}
    \label{hr}
\end{figure}
It is also important to note that $ \nabla_\mu \mathcal{T}^{\mu r}=0$, since $\gamma'=0$, thus satisfying Eq. \eqref{Tconsv}. Before analysing the Skyrme black-hole solutions in generalized teleparallel gravity, lets consider TEGR. It is also important to note that $\gamma=N\pi/2$ is also a solution for the field equation Eq. \eqref{Skyrmeeomf}, but the $U$ vanishes for even $N$.
\subsection{TEGR: $f(T)=-T-2\Lambda$}
In the case of $f(T)=-T-2\Lambda$, corresponding to the TEGR, the teleparallel equations of motions given in Eq. \eqref{teleeom1} can be written as,
\begin{align}
   &-5 r^2 l m^2 h'^2+2 r h m \left(-r m h' l'+2 l \left(r m h''+r h'
   m'+2 m h'\right)-2 \Lambda  r l^2 m\right)+h^2 \Big(2 r m l' \left(r m'+2 m\right)\nonumber \\ & +l
   \left(r^2 m'^2-4 r m \left(r m''+3 m'\right)-4 m^2\right)+4 l^2 m\Big) =\frac{16 \pi G h l^2}{e^2 r^2 }\Big(h \left(2 e^2 \kappa ^2 r^2 m+h\right)\Big)
   \nonumber
   \\
  &  \frac{h'^2}{h l}-\frac{h \left(\left(r m'+2 m\right)^2-4 l
   m\right)}{r^2 l m^2}-4 \Lambda  = \frac{16 \pi G}{e^2 r^4 m^2}\Big(h \left(2 e^2 \kappa ^2 r^2 m+h\right)\Big)
    \label{eomT1}\\
  & -\frac{h'^2}{h l}+\frac{h \left(m \left(r l' m'-2 l \left(r
   m''+2 m'\right)\right)+2 m^2 l'+r l m'^2\right)}{r l^2 m^2}-4 \Lambda  = \frac{16\pi G h^2}{e^2 r^4 m^2} \ .
   \nonumber
\end{align}
Solutions to these equations are computed as,
\begin{equation}\label{Tsol1}
\begin{split}
h(r)&=C_1-\frac{C_2}{r}+\frac{4 \pi G }{e^2r^2}-\frac{1}{3} \Lambda r^2\\
l(r)&=1\\
m(r)&=h(r)
\end{split}
\end{equation}
where $C_1$ and $C_2$ are integrating constants. Linearising the metric and comparing it to the Newtonian limit yields, $C_2=2GM$, where $M$ is the mass of the black-hole \cite{Rodrigues:2013ifa}. While, assuming the Minkowskian limit (for $\Lambda$ = 0), we get $C_1=1-8 \pi G \kappa^2$.
Thus, $h(r)$ finally becomes,
\begin{equation}
h(r):=1-8 \pi G \kappa^2-\frac{2 G M}{r}+\frac{4 \pi G \kappa^2 \lambda }{r^2}-\frac{1}{3} \Lambda r^2
\label{h(r)final}
\end{equation}
where $\lambda=1/(\kappa^2 e^2)$. Note that the black-hole solution that we obtained in Eq.\eqref{h(r)final} has been reported in the literature \cite{Canfora:2013osa}, which is found using conventional hedgehog ansatz in Riemannian geometry. A similar solution with $\lambda=0$ (without the Skyrme term) was proposed in \cite{Gibbons:1990um} to represent a global monopole within a black-hole. Furthermore, the solution with $M=\lambda=\Lambda=0$ corresponds to the Barriola-Vilenkin monopole spacetime \cite{PhysRevLett.63.341}.\\
Using Eq. \eqref{Baryoncharge}, the corresponding value of the topological charge is computed to be,
\begin{equation}
    B=0 \ .
    \nonumber
\end{equation}
The position of the Killing horizon, which marks the boundary of the black-hole region, is determined by the condition $h(r_h)=0$ \cite{Canfora:2013osa}. Further we will be taking $\Lambda=0$ as an approximation, since $\Lambda \ll 1$ (we take $\Lambda= 10^{-120} M_{Pl}^2$ \cite{Tomberg:2021ajh}) . Solving this condition for $r_h$, we get two solutions corresponding to the outer and inner horizons as,
\begin{equation}
r_{h}=\frac{G M}{1-8 \pi G \kappa^2}\left(1 \pm \sqrt{1-\frac{4 \pi (1-8 \pi G \kappa^2)}{e^2 G  M^2}}\right)
\label{horizon}
\end{equation}
Here, the upper sign gives the location of the outer horizon, while the lower sign gives the location of the inner horizon. In this paper, we will be denoting $r_h$ as the outer horizon and we assume $\kappa^2/M_{Pl}^2<1$ to keep the sanity of the model.
The mass $M$ and the horizon $r_h$ is related as
\begin{equation}
M=\frac{1}{2 G}\left((1-8 \pi G \kappa^2) r_{h}+\frac{4 \pi G }{r_{h}e^2}\right)
\nonumber
\end{equation}
For the black-hole mass $M<M_{min}$, $r_h$ becomes non-physical due to the presence of imaginary terms. The expression of  $M_{min}$ is given by
\begin{equation}
    M_{min}=\frac{2 \sqrt{\pi } \sqrt{1-8 \pi  G \kappa ^2}}{e \sqrt{G}}
    \label{Mmin}
\end{equation}
For this value of $M$, $r_h$ takes the form
\begin{equation}
    r_h=\frac{2 }{e}\sqrt{\frac{G\pi}{1-8 \pi G \kappa^2}}
    \nonumber
\end{equation}
To illustrate the behavior of $h(r)$, we plot it, in Fig. \ref{hr}, as a function of $r/r_h$ for different values of black-hole mass  M, namely $M$= $M_{min}$, $3.8 \ M_{Pl}$ and $4.0 \ M_{Pl}$.  \\
\begin{figure}[h!]
         \centering
     \begin{subfigure}{0.44\textwidth}
         \includegraphics[scale=0.64]{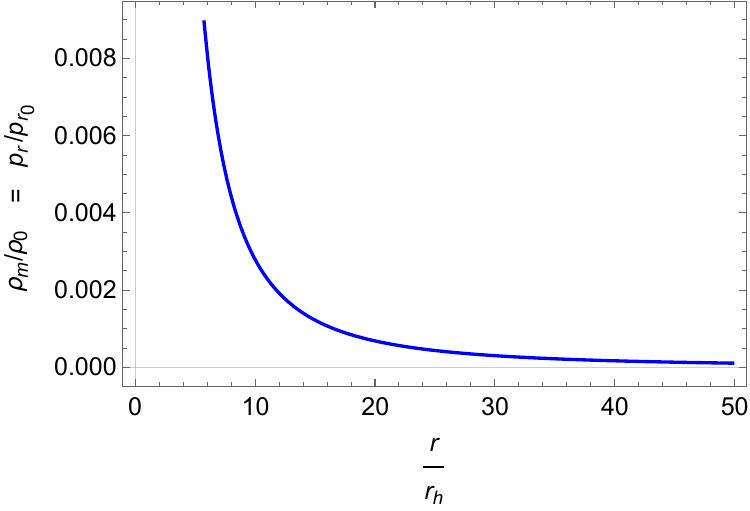}
         \caption{}
         \label{rho1}
     \end{subfigure} \hfill
     \begin{subfigure}{0.44\textwidth}
         \includegraphics[scale=0.64]{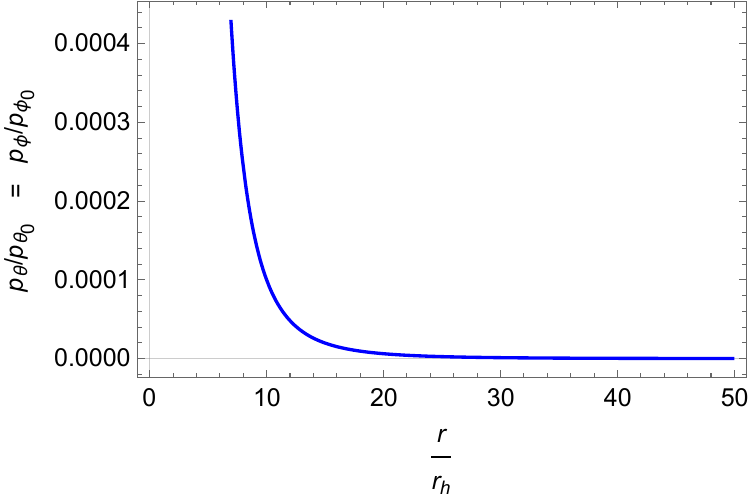}
         \caption{}
         \label{p1}
     \end{subfigure}
     \caption{The energy density $\rho_m$ equal to the radial matter pressure $p_r$ and the  $\theta$-angular pressure and $\varphi$-azimuthal pressure is shown as a function of $r/r_h$ in Fig.\eqref{rho1} and Fig. \eqref{p1} respectively. Here $\rho_0=\rho(r_h)$, $p_{r_h} = p_r(r_h)$, $p_{\theta_0} = p_\theta(r_h)$ and $p_{\varphi_0} = p_\varphi(r_h)$. Further we have chosen the coupling constant $\alpha= 8\pi G\kappa^2 = 0.1$ , $G=1 M_{Pl}^{-2}$ and $e=1$}
     \label{edpr}
     \end{figure}
Additionally, here, the energy momentum Eq. \eqref{emcase1-1}  takes the form
\begin{equation}\label{emfinal1}
\begin{split}
    \rho_m = p_r = & -\frac{1}{2 e^2 r^4 }-\frac{\kappa^2}{r^2}\\
 p_\theta = p_\varphi = &  \frac{1}{2 e^2 r^4 }
\end{split}
\end{equation}
where $\rho_m=\mathcal{T}^t{}_t$, $p_r=\mathcal{T}^r_r$, $p_\theta=\mathcal{T}^\theta{}_\theta$ and $p_\varphi=\mathcal{T}^\varphi{}_\varphi$ are the energy density, radial matter pressure and $\theta$-angular pressure and $\varphi$-angular pressure of the Skyrme field respectively. The energy density $\rho_m$ equal to the radial matter pressure $p_r$ and the  $\theta$-angular pressure and $\varphi$-azimuthal pressure is shown as a function of $r/r_h$ in Fig.\eqref{rho1} and Fig. \eqref{p1} respectively
\subsection{Weak power law $f(T)$ gravity}
\begin{figure}[h!]
     \begin{subfigure}[a]{0.44\textwidth}
         \centering
         \includegraphics[scale=0.62]{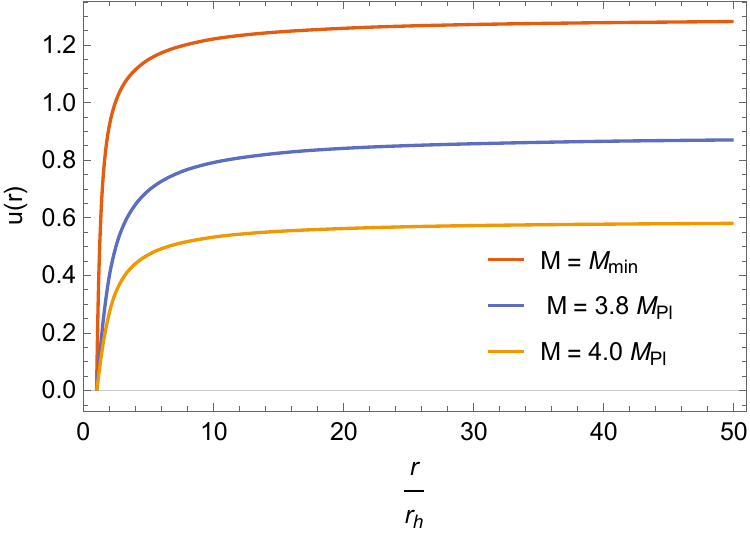}
         \caption{}
         \label{u}
     \end{subfigure} \hfill
     \begin{subfigure}[a]{0.44\textwidth}
         \centering
         \includegraphics[scale=0.62]{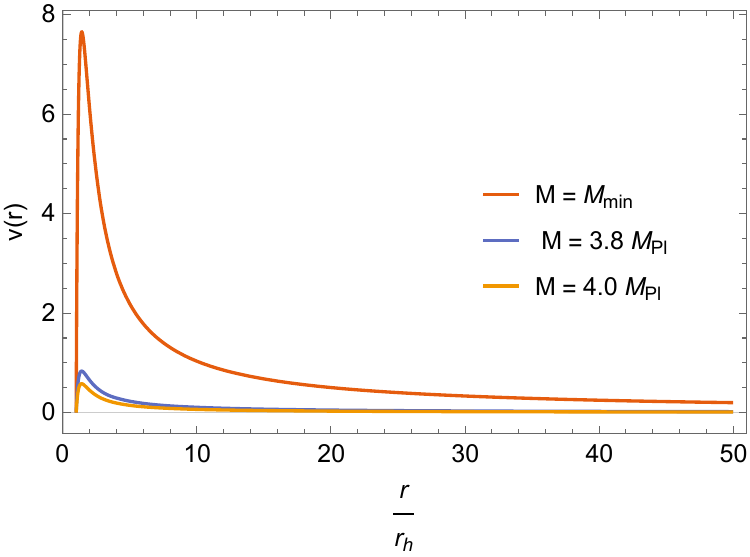}
         \caption{}
         \label{v}
     \end{subfigure}
     \caption{The functions $u(r)$ (Fig. \eqref{u}) and $v(r)$ (Fig. \eqref{v}) are shown of $r/r_h$ for different mass of black-hole $M$ namely $M$= $M_{min}$, $3.8 \ M_{Pl}$ and $4.0 \ M_{Pl}$.$M_{min}$ is computed using Eq.\eqref{Mmin} to be 0.36299 $M_{Pl}$. We have chosen $G=1 M_{Pl}^{-2}$, $\alpha= 8\pi G\kappa^2 = 0.1$,  $e=1$, $\Lambda= 10^{-120} M_{Pl}^2$ and $\tau=1M_{Pl}^{-2}$}
     \label{uvfig}
     \end{figure}
The complexity of the field equations makes it extremely challenging to obtain an analytic solution for a generic $f(T)$. Here we use perturbation theory to find approximate solutions instead of the exact analytical solutions. In this subsection, we focus on the weak power-law $f(T)$ model \cite{DeBenedictis:2016aze} given as,
\begin{equation}
    f(T)=-T-\epsilon \tau T^2 - 2 \Lambda \,
    \label{f(T)weak}
\end{equation}
where $\tau$ is the coupling constant and $0<\epsilon\ll1$ is the perturbation parameter. The assumption is made that the deviation from TEGR is small ($\epsilon\ll1$), thus only first-order terms in $\epsilon$ are considered in calculations.  We are interested in the perturbations around the metric background defined in Eq. \eqref{metricfinal} with solutions given by \eqref{Tsol1} and \eqref{h(r)final}. Hence, we choose the ansatz for $g_{tt}$ and $g_{rr}$ given as follows.
\begin{equation}
\begin{split}
    g_{tt}&=h(r)=w(r)^2+ \epsilon u(r)\\
    g_{rr}&=1/h(r)=w(r)^{-2} + \epsilon v(r)
\end{split}
\end{equation}
where
\begin{equation}
    w(r)=\sqrt{1-8 \pi G \kappa^2-\frac{2 G M}{r}+\frac{4 \pi G  \lambda}{r^2}-\frac{1}{3} \Lambda r^2 }
\end{equation}
 The functions $u(r)$ and $v(r)$ are defined as perturbations of the metric coefficients and are functions of the radial coordinate $r$. Using the $f(T)$ expansion for the weak power-law gravity up to first order in $\epsilon$ in Eq.\eqref{teleeom1} we can easily obtain the set of equations of motion.
 \begin{comment}

\end{comment}
%We also found that that the energy momentum tensor vanishes to the first order in $\epsilon$. 
These equations are solved numerically and plotted in the range $r/r_h=(1, \infty)$ in Fig. \eqref{uvfig} for different black-hole masses $M$= $M_{min}$, $3.8 \ M_{Pl}$ and $4.0 \ M_{Pl}$, where $M_{min}$ is computed using Eq.\eqref{Mmin}. Here, we have assumed that the initial conditions $u(r_h)=v(r_h)=0$. This assumption takes into account the preservation of TEGR geometry at the event horizon. 
%From Fig. \eqref{uvfig}, it is evident that $v(r)$ exhibits significantly different behavior in the vicinity of the horizon. On the other hand, $u(r)$ experiences a sudden increase near the horizon and subsequently approaches saturation as r increases.

\section{Case 2: Skyrmions with  B $\neq$ 0}
\label{sec:Bneq0}

In this section, we consider the scenario with non-trivial winding numbers for the Skyrme. We consider two cases here, TEGR where $f(T)=-T-2\Lambda$ and power law gravity where $f(T)=-T-\tau T^n-2\Lambda$( $\tau$ is a constant and $n\in \mathbb{Z}$). Moreover, unlike the discussions in the previous case, here, we will solve for the metric functions in the region of interest and look for solutions of $\gamma(r)$ such that the Skyrme equations of motions are satisfied.

Due to the complexity in solving the Skyrme field equation described by Eq. \eqref{Skyrmeeomf}, we consider two simplifying regions of interest: solutions near the event horizon denoted by $r_h$ ($(r - r_h)\ll 1$), and the far-field solution where $r\gg r_h$. To study the region close to the event horizon ($r_h$) of the black-hole, we Taylor expand Skyrme solution about $(r-r_h) \ll 1$, as, 
\begin{equation}
\label{gammaexp}
    \gamma(r)=\gamma_0 + \gamma_1(r-r_h)+\gamma_2(r-r_h)^2+ \mathcal{O}\big((r-r_h)^3\big)
    \nonumber
\end{equation}
where $\gamma_0$, $\gamma_1$ and $\gamma_2$ are constants. While, to obtain the far field solution of the Skyrme equation Eq.\eqref{Skyrmeeomf}, we assume that $r\gg r_h$. In this limit, the metric becomes flat and we can assume the Minkowski limit where $h(r), l(r),m(r)\rightarrow1$. Assuming, $\gamma(r)$ decays as $r$ goes to infinity, we choose the anzats for $\gamma(r)$, as,
\begin{equation}
\gamma(r)= f(r)/r+ \mathcal{O}(1/r ^2)
\end{equation}
where $f(r)$ is linear in $r$. Substituting the above equation in Eq. \eqref{Skyrmeeomf} and using the Minkowski limit $h(r), l(r),m(r)\rightarrow1$ and solving for $\gamma(r)$ in the far field limit, we have solved for $\gamma(r)$ as follows \cite{Herdeiro:2018daq}:
\begin{equation}
    \gamma(r)=C/r + \mathcal{O}(1/r^2)
    \nonumber
\end{equation}
where $C$ is a constant.\\
\subsection{TEGR: $f(T)=-T-2\Lambda$}
In this case, we consider the metric given by,
\begin{equation}
    d s^{2}=-h(r) d t^{2}+g(r)\left(d r^{2}+ r^{2} d \theta^{2}+ r^{2} \sin ^{2} \theta d \varphi^{2}\right)
    \label{newmetric}
\end{equation}
where $g(r)=l(r)/h(r)=m(r)/h(r)$ . Thus, we can express the equations of motion Eq \eqref{teleeom1} as,
\begin{align}
   &-\frac{4 \left(e^2 r^4 g''+2 e^2 r^3 g'-8 \pi  G r^2 \gamma'^2 \sin ^2(\gamma)-2 \pi  G \sin ^2(\gamma)+2 \pi  G \sin ^2(\gamma)
   \cos (2 \gamma)\right)}{e^2 r^4 g^2}+\frac{16 \pi  G \kappa ^2 \left(r^2 \gamma'^2+2 \sin ^2(\gamma)\right)}{r^2 g}\nonumber\\&-4 \Lambda +\frac{3
   g'^2}{g^3}=0
   \nonumber
   \\
  & \frac{4 g \left(e^2 r^3 g'+8 \pi  G r^2 \gamma'^2 \sin ^2(\gamma)-4 \pi  G \sin ^4(\gamma)\right)}{e^2}+16 \pi  G \kappa ^2 r^2 g^2 \left(r^2
   \gamma'^2+\cos (2 \gamma)-1\right)+\frac{2 r^3 g h' \left(r g'+2 g\right)}{h}\nonumber\\&+r^4 g'^2+4 \Lambda  r^4 g^3=0
    \nonumber\\
  &-\frac{2 }{g^2}\left(\frac{8 \pi  G \sin ^4(\gamma)}{e^2 r^4}+g''+\frac{g'}{r}\right)+\frac{16 \pi  G \kappa ^2 r h^2 \gamma'^2+r h'^2-2 h \left(r
   h''+h'\right)}{r h^2 g}-4 \Lambda +\frac{2 g'^2}{g^3}=0
   \nonumber
\end{align}
% Due to the complexity in solving the field equations, our focus lies solely on the near field solution of the metric and skyrme functions. 
We assume that the metric exhibits Minkowskian behavior for $r\gg r_h$.
In order to solve this, we make the assumption that the metric functions and the Skyrme function behave near the horizon ($(r-r_h)\ll1$), upto the leading order, as,
\begin{equation} \label{happro1}
\begin{split}
    h(r)&=h_0+h_1 (r-r_h)+\mathcal{O}\big(r-r_h)^2\\
    g(r)&=g_0+g_1 (r-r_h)+\mathcal{O}\big(r-r_h)^2\\
    \gamma(r)&=\gamma_0 + \gamma_1(r-r_h)+\mathcal{O}\big(r-r_h)^2
\end{split}
\end{equation}
where $h_0$, $h_1$, $g_0$, $g_1$, $\gamma_0$, $\gamma_1$ are constants. Next, we substitute Eq. \eqref{happro1}  into the teleparallel equations of motion given in Eq. \eqref{teleeom1} and solve for the near-horizon solution of the metric function and Skyrme function.
At the zeroth order of $(r-r_h)$, the teleparallel equations of motion take the following form
\begin{align}
    &e^2 r_h^2 g_0 \left(\Lambda  r_h^2 g_0-1\right)-2 \pi  G \sin ^2(\gamma_0) \left(-4 e^2 \kappa ^2 r_h^2 g_0+\cos (2 \gamma_0)-1\right)=0 \nonumber\\
   &\Lambda -\frac{4 G \pi \sin ^4(\gamma_0)}{e^2 r_h^4 g_0^2}=0\nonumber
   \end{align}
Solving the above equations for $g_0$ and $\gamma_0$, we get the following solutions.
\begin{equation}
\begin{split}
    g_0&=\frac{1}{4 e \sqrt{\pi G \Lambda} \kappa ^2  r_h^2+2 \Lambda  r_h^2} \ \ \ \text{or} \ \ \ g_0=0  \\
    \gamma_0&=\pm \sin ^{-1}\left(\frac{1}{2 \sqrt{\pi G }}\left(\sqrt{\frac{e}{2 \sqrt{\pi G } e  \kappa ^2+\sqrt{\Lambda }}}\right)\right) + 2 \pi c_1, \ \ \ \ \ c_1 \in  \mathbb{Z} 
\end{split}
\end{equation}
where $c_1$ is a constant. Here, we chose the non-zero solution of $g_0$, since $g_0=0$ leads to the solution of $B=0$. Additionally, as the negative solution of $\gamma(r)$  leads to negative topological number, we consider only the $+$ solution of $\gamma(r)$. It is important to note that $\Lambda$ has to be greater than zero in order for the above solutions to be physical. In the first order of $(r-r_h)$, we have the following equations
\begin{align}
    &\frac{4 \pi  G }{e^2}\left(\sin (\gamma_0) \left(2 e^2 g_0 \kappa ^2 r_h^2-\cos (2 \gamma_0)+1\right) (2 \gamma_1 r_h g_0 \cos (\gamma_0)-\sin (\gamma_0) (r_h
   g_1+2 g_0))\right)+g_0 r_h^2 \left(g_1 r_h+2 g_0\right)=0 \nonumber \\
   &4 \pi  G \sin ^3(\gamma_0) (\sin (\gamma_0) (r_h g_1+2 g_0)-2 \gamma_1 r_h g_0 \cos (\gamma_0))=0 \nonumber
\end{align}
Solving, we get
\begin{equation}
\begin{split}
    g_1&=-\frac{2 g_0}{r_h} \\
    \gamma_1&=0 
\end{split}
\end{equation}
Next, we will analyze the expression of the Skyrme equation in the vicinity of the horizon, represented by Eq. \eqref{Skyrmeeomf}. By solving for both the zeroth and first order terms in $(r-r_h)$, we obtain the following results.
\begin{equation}
    h_0=h_1=0  \ .
\end{equation}
  The far-field solution of $\gamma(r)$ takes the form
\begin{equation}
    \gamma(r) = 1/r + \mathcal{O}\big(1/r^2) \ .
    \nonumber
\end{equation}
Using the above obtained solutions, now  we have the near field solutions of the metric function and skyrme field as follows:
\begin{equation}
\begin{split}
     h(r)&=\mathcal{O}\big((r-r_h)^2\big)\\
   g(r)&=\frac{g_0}{3}\Bigg(1-\frac{2r}{3r_h}\Bigg)+\mathcal{O}\big((r-r_h)^2\big)\\
   \gamma(r)&=\sin ^{-1}\left(\frac{1}{2 \sqrt{\pi G }}\left(\sqrt{\frac{e}{2 \sqrt{\pi G } e  \kappa ^2+\sqrt{\Lambda }}}\right)\right) + 2 \pi c_1+\mathcal{O}\big((r-r_h)^2\big), \ \ \ \ \ c_1 \in  \mathbb{Z}
\end{split}
\end{equation}
Now, the baryon number $B$ in Eq. \eqref{Baryoncharge} becomes,
\begin{equation}
    B=\frac{1}{\pi }\left(\sin ^{-1}\left(\sigma\right)-\ \sigma \sqrt{1-{\sigma}^2}\right)+2c_1
\end{equation}
where we have  \begin{equation}
    \sigma=\frac{1}{2 \sqrt{\pi G }}\left(\sqrt{\frac{e}{2 \sqrt{\pi G } e  \kappa ^2+\sqrt{\Lambda }}}\right)
    \nonumber
\end{equation}
\begin{figure}[h!]
    \centering
    \includegraphics[scale=0.64]{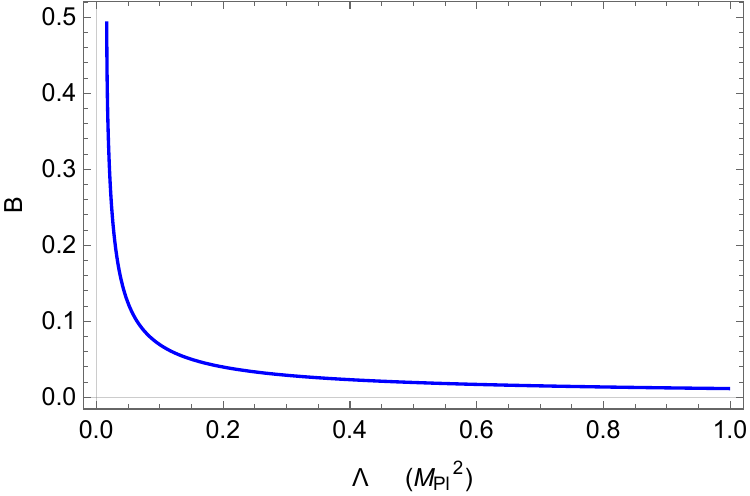}
    \caption{The variation of B with respect to the cosmological constant $\Lambda$ in the case of TEGR for $8\pi G \kappa^2=0.1$, $G=1{M_{Pl}}^{-2}$, $e=1$, $\tau=1/2 M_{Pl}^{-2} $ and $c_1=0$}
    \label{figBTEGR}
\end{figure}
 It is evident that the cosmological constant $\Lambda$ has be positive for the existence of Skyrme in TEGR. And, Fig. \eqref{figBTEGR} shows the variation of $B$ with respect to $\Lambda$. Further $B$ tends to zero (or $c_1$), as $\Lambda$ goes to infinity. This behaviour matches with the results in Einstein-Skyrme system \cite{Brihaye:2005an}.
\subsection{Power law $f(T)$ gravity}
In this section, we aim to find solutions for the metric and Skyrme fields within the framework of the power law $f(T)$ gravity model, represented by the equation \cite{DeBenedictis:2016aze}:
\begin{equation}
f(T)=-T-\tau T^n-2\Lambda
\label{powerlaw}
\end{equation}
As the field equations become increasingly intricate for $n>2$, we will focus on the case where $n=2$ for the sake of simplicity. In this scenario, we will consider the metric given in Eq. \eqref{newmetric} and, again, since the field equations are complex to solve, we focus exclusively on the near field solution of the metric and Skyrme functions and also assume that the metric becomes Minkowski for large values of $r$. Similar to the previous case of TEGR, we assume that the behaviour of metric and Skyrme function, near the horizon ($(r-r_h)\ll1$), is as given in Eq. \eqref{happro1}. Substituting Eq. \eqref{powerlaw} and Eq. \eqref{happro1} in the teleparallel equations of motion Eq. \eqref{teleeom1}, the teleparallel field equations takes the following form,
\begin{align}
   &e^2 \left(g_0^2 \Lambda  r_h^4-g_0 r_h^2+2 \tau \right)-2 \pi  G \sin ^2\left(\gamma _0\right) \left(\cos \left(2 \gamma _0\right)-4 e^2 g_0 \kappa ^2 r_h^2-1\right)=0 \nonumber\\
   &\frac{\Lambda }{2}-\frac{2 G\pi\sin ^4\left(\gamma _0\right)}{2 e^2 g_0^2 r_h^4}-\frac{\tau }{g_0^2 r_h^4}=0 \nonumber
\end{align}
On solving this, we get
\begin{equation}\label{fteq}
\begin{split}
    {g_0}_{\pm}=&-\frac{\Lambda \pm 2 \sqrt{\pi } \sqrt{e^2 G \kappa ^4 \Lambda  \left(32 \pi  e^2 G \kappa ^4 \tau -8 \Lambda  \tau +1\right)}}{8 \pi  e^2 G \kappa ^4 \Lambda  r_h^2-2 \Lambda ^2
   r_h^2} \ \ \ \text{or} \ \ \ g_0=0  \\
    \gamma_0=& \sin ^{-1}\left(\sqrt{e} \left(\frac{g_{0\pm}^2 \Lambda  r_h^4-2 \tau }{4\pi G}\right)^\frac{1}{4}\right)+2 \pi  c_1, \ \ \ \ c_1 \in \mathbb{Z} 
\end{split}
\end{equation}
We choose the non-zero solution of $g_0$ for the reason mentioned in the previous section. 
Now, to the first order of $(r-r_h)$, the teleparallel equations takes the following form,
\begin{align}
    &4 \pi  G \sin (\gamma_0) \left(2 e^2 \kappa ^2 r_h^2 g_0-\cos (2 \gamma_0)+1\right) (2 \gamma_1 r_h g_0 \cos (\gamma_0)-\sin (\gamma_0) (r_h
   g_1+2 g_0))+e^2 \left(r_h^2 g_0-4 \tau \right) (r_h g_1+2 g_0)=0 \nonumber\\
    &2 (r_h g_1+2 g_0) \left(e^2 \tau +2 \pi  G \sin ^4(\gamma_0)\right)-8 \pi  G \gamma_1 r_h g_0 \sin ^3(\gamma_0) \cos (\gamma_0)=0 \nonumber
\end{align}
After solving this system of equations, we obtain the following results,
\begin{equation}
\begin{split}
\gamma_1&=0 \\
    g_1&=-\frac{2{g_0}_\pm}{r_h} 
\end{split}
\end{equation}
Now, let us examine the near-horizon form of the Skyrme equation, as given by Eq. \eqref{Skyrmeeomf}. Solving for zeroth and first order in $(r-r_h)$, we get
\begin{equation}
    h_0=h_1=0 
\end{equation}
Thus, the near field solutions of the metric function and skyrme field are given as,
\begin{equation}\label{fteq}
\begin{split}
    h(r)&=\mathcal{O}\big((r-r_h)^2\big)\\
   g(r)&=\frac{{g_0}_{\pm}}{3}\Bigg(1-\frac{2r}{3r_h}\Bigg)+\mathcal{O}\big((r-r_h)^2\big)\\
   \gamma(r)&= \sin ^{-1}\left(\alpha_\pm \right)+2 \pi c_1 +\mathcal{O}\big((r-r_h)^2\big), \ \ \ \ c_1 \in \mathbb{Z}
\end{split}
\end{equation}
where we have
\begin{equation}
    \alpha_\pm = \sqrt{e} \left(\frac{{g_0}_{\pm}^2 \Lambda  r_h^4-2 \tau }{4\pi G}\right)^\frac{1}{4}. 
    \nonumber
\end{equation}
\begin{figure}[h!]
    \centering
    \includegraphics[scale=0.26]{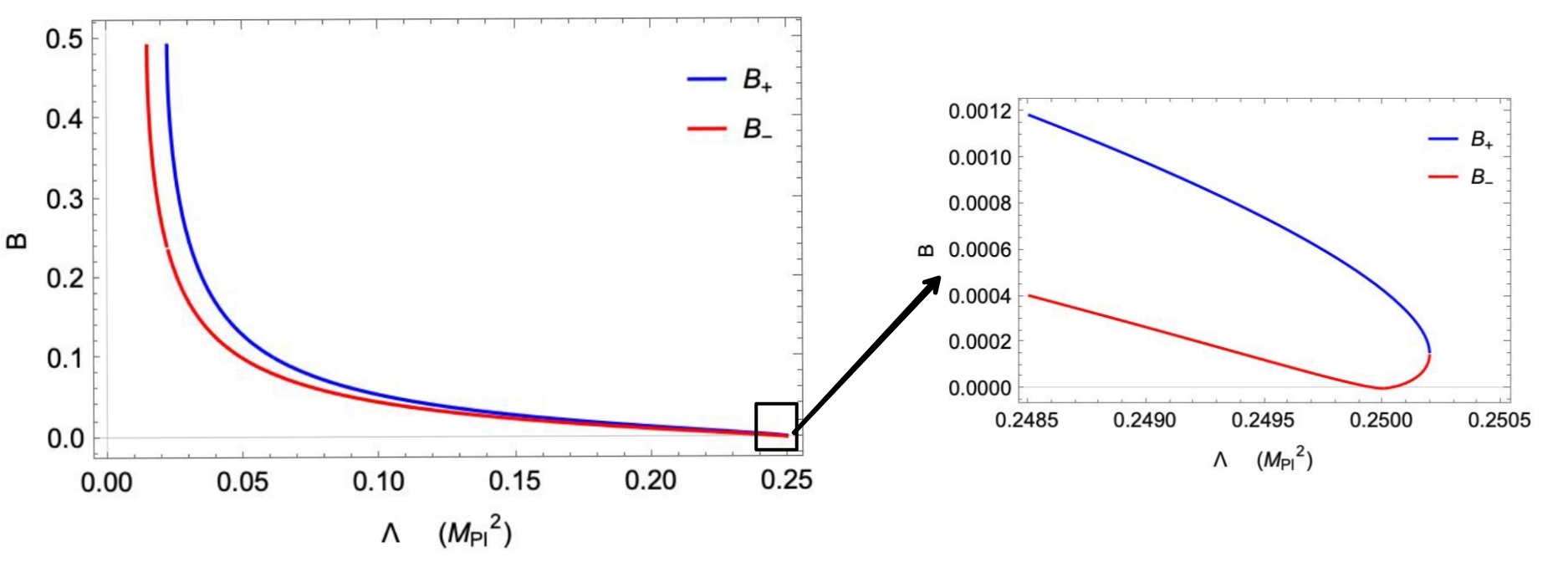}
    \caption{The variation of B with respect to the cosmological constant $\Lambda$ for $8\pi G\kappa^2=0.1$, $G=1{M_{Pl}}^2$, $e=1$, $\tau=1/2 M_{Pl}^{-2} $ and $c_1=0$}
    \label{figBfT}
\end{figure}
And the far-field solution of $\gamma(r)$ becomes,
\begin{equation}
    \gamma(r) = 1/r + \mathcal{O}\big(1/r^2) \ .
    \nonumber
\end{equation}
Further, the reality condition on ${g_0}_{\pm}$ and $\gamma_0$ given in Eq. \eqref{fteq} (we name it as condition-1) is given by,
\begin{equation}\label{cond1}
    \Lambda \leq \frac{32 \pi  e^2 G \kappa ^4 \tau  +1}{8 \tau}
\end{equation}
Now we can use Eq. \eqref{Baryoncharge} to find the expression of the topological number $B$, which is obtained as,
\begin{align}
    B_{\pm}&=\frac{1}{\pi }\left(\sin ^{-1}\left(\alpha_\pm\right)-\alpha_\pm \sqrt{1-{\alpha_\pm}^2}\right)+2c_1, \ \ \ \  c_1 \in \mathbb{Z}
   \label{B2}
\end{align}
$B_{+}$ and $B_{-}$ are the topological numbers corresponding to the two different solutions, which depends on the cosmological constant $\Lambda$, but independent of the event horizon $r_h$. Further both solutions coincided when $\kappa\rightarrow 0$. The reality condition of $B_{\pm}$ then becomes,
\begin{figure}[h!]
    \centering
    \begin{subfigure}[b]{0.45\textwidth}
    \centering
    \includegraphics[scale=0.63]{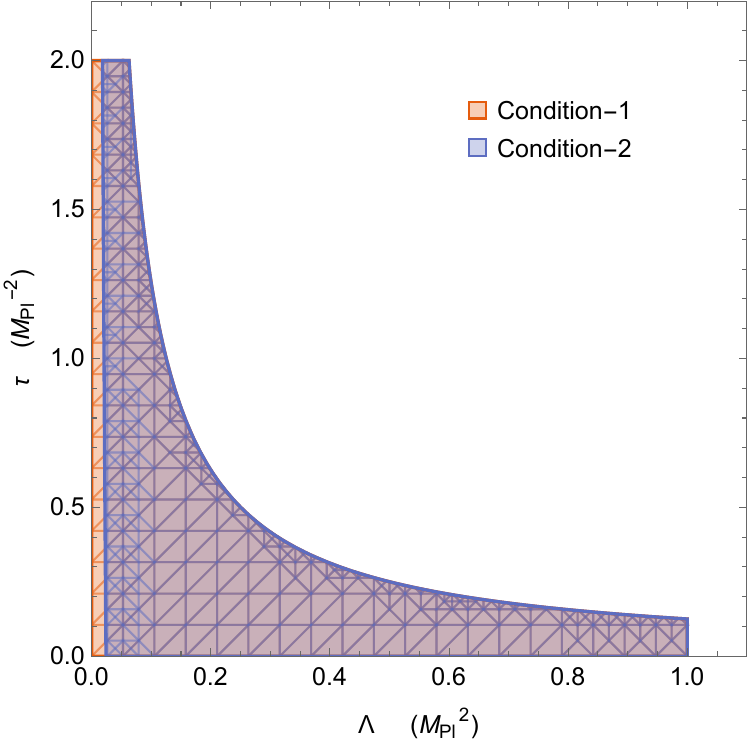}
    \caption{}
    \label{fig:tauconditionsa}
\end{subfigure}
\hfill
\begin{subfigure}[b]{0.45\textwidth}
    \centering
    \includegraphics[scale=0.60]{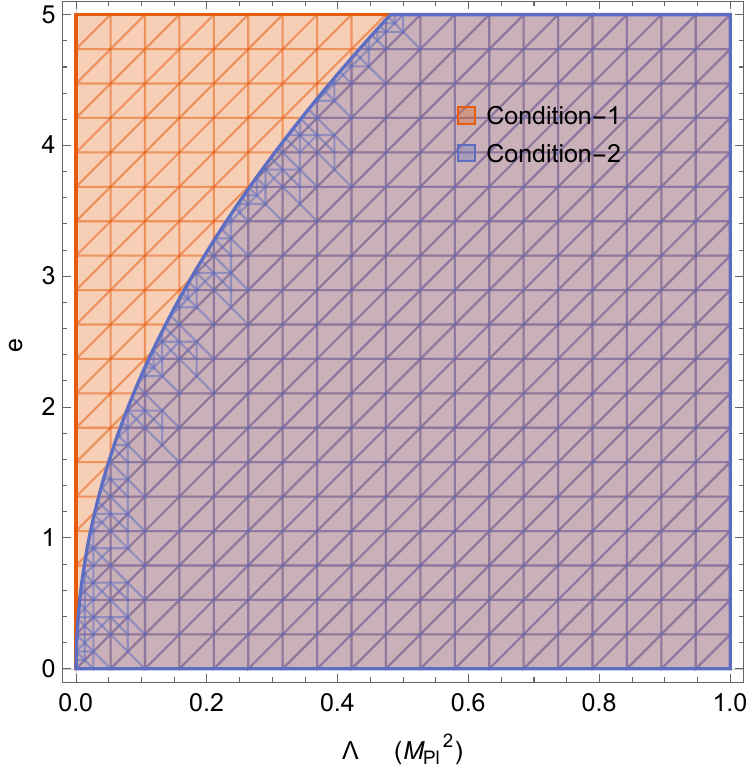}
    \caption{}
    \label{fig:tauconditionsb}
     \end{subfigure}
     \caption{Here we have shown the parameter space that satisfies condition-1 and condition-2 given in Eq.\ref{cond1} and Eq.\ref{cond2}  respectively as a function of $\Lambda$ and $\tau$ \eqref{fig:tauconditionsa} for $8\pi G \kappa^2=0.1$, $G=1{M_{Pl}}^{-2}$ and $e=1$ and as a function of $\Lambda$ and $e$ \eqref{fig:tauconditionsb} for $8\pi G \kappa^2=0.1$, $G=1{M_{Pl}}^{-2}$ and $\tau=0.01$. The overlap region represents the region where $g_{0\pm}$, $\gamma_0$ and $B_{\pm}$ are real. This also shows that there exists minimum and maximum for the cosmological constant ($\Lambda_{min}$ and $\Lambda_{max}$), for given $\tau$ in \eqref{fig:tauconditionsa}.}
\end{figure}
\begin{equation}\label{cond2}
    \alpha_\pm\leq1 \implies e^2 \left(\frac{\Lambda  \left(2 \sqrt{\pi } \sqrt{e^2 G \kappa ^4 \Lambda   \left(32 \pi  e^2 G \kappa ^4 \tau -8 \Lambda  \tau +1\right)}+\Lambda 
   \right)^2}{\left(8 \pi  e^2 G \kappa ^4 \Lambda -2 \Lambda ^2 \right)^2}-2 \tau \right)\leq 4 \pi  G
\end{equation}
In Fig. \eqref{figBfT}, we have shown the variation of $B$ with respect to $\Lambda$. We can see that there is a $\Lambda_{max}$ above which $B$ does not exist. Further, there is also a $\Lambda_{min}$ below which $B$ does not exist. In the limit as $\tau$ approaches zero, the geometry transforms into TEGR. In this limit, $\Lambda_{max}$ approaches infinity and $\Lambda_{min}$ approaches zero. Consequently, we recover the result obtained from TEGR, thereby ensuring consistency. In Fig.\eqref{fig:tauconditionsa}, we have plotted dependence of the lower and upper limits of $\Lambda$ on $\tau$ parameter. Whereas in Fig. \eqref{fig:tauconditionsb}, we have shown the parameter space that satisfies condition-1 and condition-2 given in Eq.\ref{cond1} and Eq.\ref{cond2} respectively, as a function of $e$ and $\Lambda$ for $8\pi G \kappa^2=0.1$, $G=1{M_{Pl}}^{-2}$ and $\tau=0.01$. This is an interesting new result which states that, unless $e \to 0$ the cosmological constant $\Lambda$ cannot vanish. Remembering that $e^2$, in Eq.\ref{eq:skyrme}, is the dimensionless coupling constant that determines the size of the soliton with respect to $\kappa^2$, the lower bound on the energy of the soliton ($E \geq \frac{6 \pi^2 \kappa}{e}|B|$) becomes very large.  

\section{Results and Discussions}
\label{sec:results}
In this study, we explored the intriguing field of Skyrmions within the framework of teleparallel gravity, focusing on two distinct scenarios characterized by either a zero or a non-zero baryon number ($B=0$ and $B\neq0$). Our analysis involved a detailed computation of the Skyrme solutions and their associated metric modifications in the contexts of both TEGR ($f(T) = -T - 2\Lambda$) and power law gravity. For the generalized teleparallel $f(T)$ gravity, we employed the power gravity model as outlined in Eq.\eqref{f(T)weak}, utilizing perturbative techniques to derive essential corrections. Specifically, for the $B\neq0$ scenario, we conducted an extensive calculation of the Skyrme solution and metric perturbations, taking into account both the near-horizon limit ($(r - r_h)\ll 1$) and the far limit ($r\gg r_h$), thus offering a comprehensive view of the behavior in these distinct regimes.\\
Our findings revealed that the TEGR solutions for the $B=0$ case are in agreement with those of the established Einstein-Skyrme system. Interestingly, we observed that in such cases, the Skyrme function $\gamma(r)$ remains unaffected by variations in the cosmological constants and Skyrme parameters. This contrasts with the $B\neq0$ scenario, where the Skyrme function demonstrates a clear dependency on the cosmological constant $\Lambda$. Moreover, in the $f(T) = -T - 2\Lambda$ framework with $B\neq0$, the Skyrme solution's physical plausibility is contingent on a positive cosmological constant, aligning with results previously identified in the Einstein-Skyrme system, as referenced in \cite{Brihaye:2005an}.\\
In a more nuanced development within the power law gravity model, as detailed in Eq.\eqref{powerlaw}, the Skyrme solution and baryon number are found to be influenced not only by the cosmological constant but also by the introduction of specific lower ($\Lambda_{min}$) and upper ($\Lambda_{max}$) bounds. These bounds are critical in ensuring the physical feasibility of the Skyrmion, stipulating that the cosmological constant must lie within the specified range $\Lambda_{min} < \Lambda < \Lambda_{max}$. This introduces a new layer of complexity to the model, as these bounds are not merely arbitrary limits but are closely tied to the physical characteristics of the system. The results illustrating these findings are depicted in Fig.\eqref{figBfT}. Furthermore, we analyzed how the determination of the lower and upper limits of $\Lambda$ is intricately linked to the $\tau$ parameter. This relationship is clearly demonstrated in Fig.\eqref{fig:tauconditionsa}, where the dependency of these bounds on the $\tau$ value is graphically represented, offering a deeper insight into the dynamics of the power law gravity model in relation to the Skyrme solutions.

Moreover, Fig.\eqref{fig:tauconditionsb}, shows the relation of the Skyrme size to the the cosmological constant. This result is very interesting and shows that for vanishing cosmological constant, the energy of the Skyrme system has to be very large.

\begin{acknowledgments}
The authors would like to thank Prof. Jutta Kunz and
Prof. Eugen Radu for their helpful comments and discussions on this work. M.T.A. acknowledges financial support of DST through INSPIRE Faculty grant [DST/INSPIRE/04/2019/002507].
\end{acknowledgments}
\appendix
\section{Components of the Skyrme field and energy momentum tensor}
\label{A1}
The respective components of the Skyrme field $Q_\mu$ can be written, using Eq. \eqref{U}, as
\begin{eqnarray}
    Q_t &=& 0
\nonumber\\
Q_r &=& \dis \begin{bmatrix}
 i\cos \theta \gamma'  & \sin \theta \gamma' \left(i \cos \varphi + \sin \varphi\right) \\ 
  \sin \theta \gamma' \left(i \cos \varphi - \sin \varphi\right) &  - i\cos \theta \gamma'
\end{bmatrix} 
\nonumber \\
    Q_\theta &=& \dis  \begin{bmatrix}
  -i \cos \gamma \sin \gamma \sin \theta   & \sin \gamma \left(\cos \phi - i \sin \phi \right)\left(i\cos \theta \cos \gamma + \sin \gamma \right) \\ 
  \sin \gamma \left(\cos \phi + i \sin \phi \right)\left(i\cos \theta \cos \gamma - \sin \gamma \right) &   i \cos \gamma \sin \gamma \sin \theta 
\end{bmatrix} 
\label{Q3} \\
    Q_\varphi &=& \dis \begin{bmatrix}
  i \sin^2 \theta \sin^2 \gamma  &  \sin \gamma \sin \theta \left(\cos \phi - i \sin \phi \right)\left(\cos \gamma +i\cos \theta  \sin \gamma \right) \\ 
 - \sin \gamma \sin \theta \left(\cos \phi + i \sin \phi \right)\left(\cos \gamma -i\cos \theta  \sin \gamma \right)  &  -i \sin^2 \theta \sin^2 \gamma
\end{bmatrix} 
\nonumber 
\end{eqnarray}
Using this, the non-zero components of energy momentum tensor $T_{\mu \nu}$ of the Skyrme field given in Eq. \eqref{emtensor} becomes,
\begin{equation}\label{T11}
\begin{split}
    \mathcal{T}^{tt}= &\frac{1}{2 e^2 r^4 p_1 p_2
   p_3}\Bigg( e^2 \kappa ^2 r^2 \left(r^2 p_2p_3 {\gamma'}^2+p_1 \sin ^2\gamma (p_2+p_3)\right)+h
   \left(r^2 {\gamma'}^2 \sin ^2\gamma (p_2+p_3)+p_1 \sin ^4\gamma\right)\Bigg) \\
\mathcal{T}^{rr} = & \frac{1}{16 e^2 r^4 p_1^2 p_2 p_3}h^2 \left(4 e^2 \kappa ^2 r^2 \left(2 r^2 p_2 p_3 {\gamma'}^2-2 p_1 \sin ^2\gamma
   (p_2+p_3)\right)+4 h \sin ^2\gamma \left(2 r^2 {\gamma'}^2 (p_2+p_3)-2 p_1 \sin
   ^2\gamma\right)\right)\\
\mathcal{T}^{\theta\theta} = & \frac{1}{2 e^2
   r^6 p_1 p_2^2 p_3}h^2 \left(h \sin ^2\gamma \left(r^2 {\gamma'}^2 (p_3-p_2)+p_1 \sin ^2\gamma\right)-\frac{1}{2} e^2
   \kappa ^2 r^2 \left(2 r^2 p_2 p_3 {\gamma'}^2+2 p_1 \sin ^2\gamma (p_2-p_3)\right)\right)\\
\mathcal{T}^{\varphi\varphi} =&  \frac{1}{16 e^2 r^6 p_1 p_2 p_3^2}h^2 \csc ^2(\theta ) \left(8 h \sin ^2\gamma \left(r^2 {\gamma'}^2 (p_2-p_3)+p_1 \sin
   ^2\gamma\right)-4 e^2 \kappa ^2 r^2 \left(2 r^2 p_2 p_3 {\gamma'}^2-2 p_1 \sin ^2\gamma
   (p_2-p_3)\right)\right)  
\end{split}
\end{equation}
Further, demanding the conservation of energy momentum,
\begin{equation}
    \nabla_\mu \mathcal{T}^{\mu \nu}=0 \ ,
    \label{Tconsv}
\end{equation}
we get,
\begin{equation} \label{Tconsv1}
    \begin{split}
    \nabla_\mu \mathcal{T}^{\mu t} = & 0\\
        \nabla_\mu \mathcal{T}^{\mu r} = & \frac{h^2 \gamma'}{2 e^2 r^4 p_1^3 p_2^2 p_3^2}\eta(r) \\
    \nabla_\mu \mathcal{T}^{\mu \theta}=& -\frac{1}{e^2 r^4
   p_1 p_2^2 p_3}h^2 \cot (\theta ) \sin ^2\gamma (p_2-p_3) \left(e^2 \kappa ^2 p_1+h \gamma'^2\right)\\
   \nabla_\mu \mathcal{T}^{\mu \varphi} = & 0
    \end{split}
    \end{equation}
where we have
\begin{align}
   \eta(r)&= \Big(r^2 p_2 p_3 (p_1 (p_3 \gamma' (e^2 \kappa ^2 r^2 p_2'+2
   h' \sin ^2\gamma)+p_2 (\gamma' (e^2 \kappa ^2 r^2 p_3'+2 h' \sin ^2\gamma)+2 e^2
   \kappa ^2 r p_3 (r \gamma''+2 \gamma'))) \nonumber \\ &-e^2 \kappa ^2 r^2 p_2 p_3 \gamma'p_1'-e^2 \kappa ^2 p_1^2 \sin (2 \gamma) (p_2+p_3))- h \sin \gamma(r^2 -p_2  p_3 \gamma' \sin \gamma p_1' (p_2+p_3)+r^2 p_1 (-p_3^2 \gamma' \sin \gamma p_2' +p_2 p_3 \times \nonumber \\ &(\gamma'  \sin \gamma(p_2'+p_3')+2 p_3
   (\gamma'' \sin \gamma+\gamma'^2 \cos \gamma))+p_2^2 (2 p_3 (\gamma'' \sin
   \gamma+\gamma'^2 \cos \gamma)-\gamma' \sin \gamma p_3')) -4 p_1^2 p_2
   p_3 \sin ^2\gamma \cos \gamma )\Big)
   \nonumber
\end{align}

\section{Teleparallel equations of motion}
\label{A2}
It is important to note that the above metric is spherically symmetric \cite{Ioannidou:2006mg} and is of the form,
\begin{equation}
d s^{2}=g_{A B}(y) d y^{A} d y^{B}+\beta(y)^{2} \gamma_{a b}(z) d z^{a} d z^{b}
\nonumber
\label{sphmetric}
\end{equation}
where $y^A=\{t,r\}$ and $g_{AB}$ are the coordinates and the metric on two-dimensional Lorentzian $M^2$ respectively, while $z^a=\{\theta,\varphi\}$ and $\gamma_{ab}$ are the coordinates and the metric on the two-dimensional unit sphere $S^2$.\\
We choose the tetrad that satisfies the metric as,  
\begin{equation}
\label{diagtetard}
\left.h_{\mu}{ }^{a}\right|_{\operatorname{diag}}=\operatorname{diag}\left(\sqrt{h(r)},\sqrt{l(r)/h(r)},r\sqrt{m(r)/h(r)},r\operatorname{sin(\theta)}\sqrt{l(r)/h(r)}\right)
\nonumber
\end{equation}
While this tetrad generates the metric in Eq.\eqref{metricfinal}, it is not suitable for reducing the algebraic expression of $f(T)$ to the teleparallel linear form. Therefore, we follow the approach developed in literatures \cite{Tamanini:2012hg,Maluf:2007qq, Maluf:2008ug, Boehmer:2012uyw}, where a local Lorentz transformation in the tangent space is performed to construct a good set of non-diagonal tetrads. These transformations, in the tangent space (with flat metric $\eta_{a b}$), are given by, $h_{\mu}^{a} \mapsto \Lambda_{b}^{a} h_{\mu}^{b}$, where $\Lambda_{b}^{a}$ satisfies the condition $\eta_{a c} \Lambda_{b}^{a} \Lambda_{d}^{c}=\eta_{b d}$. By using these non-diagonal "good" tetrads, we obtain the desired algebraic teleparallel equations of motion for the Skyrme black-hole system.To construct our tetrads, we choose an ansatz that is adapted to stationary observers with a four-velocity $\left( u^0,0,0,0\right)$. This is achieved by setting the degree of freedom to be $h_0^\lambda = u^\lambda$, while the other components $ h_1^\lambda$, $ h_2^\lambda$, and $h_3^\lambda$ are aligned with the $\widehat{x}$, $\widehat{y}$, and $\widehat{z}$ Cartesian directions. With this ansatz, we obtain the tetrad field as,
\begin{equation}
h_{\mu}^{a}=\left[\begin{array}{cccc}
\sqrt{h(r)} & 0 & 0 & 0 \\
0 & \sqrt{l(r)/h(r)} \sin \theta \cos \phi & \ \ r \sqrt{m(r)/h(r)} \cos \theta \cos \phi & \ \ -r \sqrt{m(r)/h(r)} \sin \theta \sin \phi \\
0 & \sqrt{l(r)/h(r)} \sin \theta \sin \phi & \ \ r \sqrt{m(r)/h(r)} \cos \theta \sin \phi & r \sqrt{m(r)/h(r)} \sin \theta \cos \phi \\
0 & \sqrt{l(r)/h(r)} \cos \theta & \ -r \sqrt{m(r)/h(r)} \sin \theta & 0
\end{array}\right]
\label{tetradnew}
\end{equation}
A spherically symmetric teleparallel geometry is defined consistently by this tetrad. Using this, the non-zero components of the  Wietzenbock connection are computed to be ,
\begin{align}
\Gamma^{t}{}_{tr} =& \frac{h'(r)}{2 h(r)} & 
\Gamma^{r}{}_{rr} =& \frac{1}{2} \left(\frac{l'(r)}{l(r)}-\frac{h'(r)}{h(r)}\right) &
 \Gamma^{r}{}_{\theta\theta}= & -r \sqrt{\frac{m(r)}{l(r)}} \nonumber \\
\Gamma^{r}{}_{\varphi\varphi} = & -r \sqrt{\frac{m(r)}{l(r)}}\sin ^2(\theta ) &
 \Gamma^{\theta}{}_{r\theta} = &  \Gamma^{\varphi}{}_{r\varphi} = \frac{1}{r}\sqrt{\frac{l(r)}{ m(r)}} &
 \Gamma^{\theta}{}_{\theta r} =  &  \Gamma^{\varphi}{}_{\varphi r} = -\frac{h'(r)}{2 h(r)}+\frac{m'(r)}{2 m(r)}+\frac{1}{r} \label{wbconn} \\
 \Gamma^{\theta}{}_{\varphi\varphi} =  & -\sin\theta\cos\theta &
\Gamma^{\varphi}{}_{\theta\varphi} = &  \Gamma^{\varphi}{}_{\varphi \theta}= \cot \theta \nonumber \ .
\end{align}
Further, the non-zero components of the torsion and contorsion tensor becomes,
\begin{align}
   T^{t}{}_{rt}=& \frac{h'(r)}{2 h(r)}   \ \ \ \ \ \ \ \ \ \ 
    T^{\theta}{}_{\theta r}=  T^{\varphi}{}_{\varphi r}= \frac{h'(r)}{2 h(r)}+\frac{1}{r}\sqrt{\frac{l(r)}{ m(r)}}-\frac{m'(r)}{2
   m(r)}-\frac{1}{r} 
   \label{torsion} 
   \end{align}
  \begin{align}
    K^{\theta}{}_{r\theta} = & K^{\varphi}{}_{r\varphi}=  \frac{1}{2} \left(\frac{h'(r)}{h(r)}+\frac{2}{r}\sqrt{\frac{l(r)}{ m(r)}}-\frac{m'(r)}{m(r)}-\frac{2}{r}\right)  \ \ \ \ \ \ \ \ \ \ 
    K^{r}{}_{tt}=  -\frac{h(r) h'(r)}{2 l(r)} \nonumber \\
     K^{r}{}_{\theta\theta}=& -r\sqrt{\frac{m(r)}{\l(r)}}-\frac{r^2}{2}\frac{h'(r)m(r)}{h(r)l(r)}+\frac{r}{2l(r)}\left(r
   m'(r)+2 m(r)\right)  \ \ \ \ \ \ \ \ \ \ 
     K^{t}{}_{rt}=-\frac{h'(r)}{2 h(r)} \label{contor} \\
     K^{r}{}_{\varphi\varphi}=& r \sin ^2\theta \left(-r\sqrt{\frac{m(r)}{\l(r)}}-\frac{r^2}{2}\frac{h'(r)m(r)}{h(r)l(r)}+\frac{r}{2l(r)}\left(r m'(r)+2 m(r)\right)\right) \nonumber \ .
     \end{align}
Using the above equations (Eq. \eqref{wbconn}  Eq. \eqref{torsion} and \eqref{contor}), the components of the dual torsion tensor can be written as, 
   \begin{align}
    S_{t}{}^{tr}=&\frac{h(r)}{l(r}\left(\frac{h'(r)}{2 h(r)}+\frac{1}{r}\sqrt{\frac{l(r)}{ m(r)}}-\frac{m'(r)}{2
   m(r)}-\frac{1}{r}\right) \nonumber \\
   S_{\theta}{}^{\theta r}= & S_{\varphi}{}^{\varphi r} = \frac{1}{4 r
   l(r) m(r)}\left(h(r) \left(-2 h(r) \sqrt{\frac{l(r)}{h(r)}} \sqrt{\frac{m(r)}{h(r)}}+r m'(r)+2 m(r)\right) \right) \ .
   \label{dualtor}
\end{align}
 And, the torsion scalar can be derived as,
\begin{equation}
    T=\frac{1}{2 r^2 l m^3}\left(\frac{r^2 m^3 h'^2}{h}+4 h^3 \sqrt{\frac{l}{h}} \left(\frac{m}{h}\right)^{3/2} \left(r m'+2
   m\right)-h m \left(4 l m+\left(r m'+2 m\right)^2\right)\right)
   \label{TScalar}
\end{equation}
In the Minkowski limit ($h(r) \rightarrow 1$, $l(r) \rightarrow 1$, $m(r) \rightarrow 1$), it is straightforward to see that the torsion scalar vanishes i.e $T=0$. This is expected because the torsion scalar is a measure of the geometry's deviation from flat Minkowski spacetime. There is no such deviation in the Minkowski limit, and hence the torsion scalar vanishes.
%While there are multiple tetrads that could be used to get the desired metric, not all of them yield torsion scalars that consistently vanish in the Minkowski space-time limit. As such, it is important to select the "good" tetrad that satisfies this condition.\\
Substituting Eq.\eqref{wbconn}-\eqref{TScalar} and Eq.\eqref{T11} in Eq.\eqref{eom1}, we get the teleparallel equations of motion as,
\begin{align}
& \frac{1}{r m}\Bigg(f_T l' \Big(r m h'-h (r m'+2 m)\Big)\Bigg)-\frac{1}{r^2 h m^2}\Bigg(2 l
   \Big(-r^2 f_T m^2 h'^2+r h m \big(r f_T h' m'+m \big(r f_{\text{TT}} h' T'+f_T (r
   h''+2 h')\big)\big) \nonumber \\ & +h^3 \sqrt{\frac{l}{h}} \sqrt{\frac{m}{h}} \left(r f_T m'+2 m
   \left(r f_{\text{TT}} T'+f_T\right)\right)-h^2 m \left(r \left(r f_{\text{TT}} m' T'+f_T \left(r m''+4
   m'\right)\right)+2 m \left(r f_{\text{TT}} T'+f_T\right)\right)\Big)\Bigg)+f l^2\nonumber\\ & +\frac{\pi G}{ e^2 r^4 l m^2}\Big(2 r^2 m \gamma'^2 \left(-e^2 \kappa ^2 r^2 m+\cos (2 \gamma)-1\right)+l \sin ^2(\gamma) \left(-4 e^2 \kappa ^2 r^2 m+\cos (2
   \gamma)-1\right)\Big)=0 \nonumber \\
    &-r^2 f_T m^2 h'^2-2 f_T h^3 \sqrt{\frac{l}{h}} \sqrt{\frac{m}{h}} \Big(r m'+2
   m\Big)+f r^2 h l m^2+f_T h^2 \left(r m'+2 m\right)^2+\frac{\pi G}{ e^2 r^4 l m^2}\Big(2 r^2 m \gamma'^2 \big(-e^2 \kappa ^2 r^2 m\nonumber\\&+\cos (2 \gamma)-1\big)+l \sin ^2(\gamma) \left(-4 e^2 \kappa ^2 r^2 m+\cos (2
   \gamma)-1\right)\Big)=0  \label{teleeom1}\\ 
   &-r f_T h m l' \big(r m'+2 m\big)+2 h l \Big(-2 r f_T h
   \sqrt{\frac{l}{h}} \sqrt{\frac{m}{h}} m'+m \Big(r \big(r f_{\text{TT}} m' T'+f_T (r m''+4
   m')\big) \nonumber \\ &-2 h \sqrt{\frac{l}{h}} \sqrt{\frac{m}{h}} \big(r f_{\text{TT}} T'+2
   f_T\big)\Big)+2 m^2 \big(r f_{\text{TT}} T'+f_T\big)\Big)+2 l^2 m \big(2 f_T h+f r^2
   m\big)+ 2\pi G \left(\frac{\sin ^4(\gamma(r))}{e^2 r^4 m(r)^2}-\frac{\kappa ^2 \gamma'(r)^2}{l(r)}\right) =0
    \nonumber
\end{align}
where $f_T=\frac{\partial f}{\partial T}$ and $f_{TT}=\frac{\partial^2 f }{\partial T^2}$. 
\bibliography{ref}

\end{document}